# A unified approach to describe optical pulse generation by propagation of periodically phase-modulated CW laser light

**Víctor Torres-Company and Jesús Lancis**
Departament de Ciències Experimentals, Universitat Jaume I, E12080 Castelló, Spain
lancis@exp.uji.es

**Pedro Andrés**
Departamento de Óptica, Universitat de València, E46100 Burjassot, Spain

**Abstract:** The analysis of optical pulse generation by phase modulation of narrowband continuous-wave light, and subsequent propagation through a group-delay-dispersion circuit, is usually performed in terms of the so-called bunching parameter. This heuristic approach does not provide theoretical support for the electrooptic flat-top-pulse generation reported recently. Here, we perform a waveform synthesis in terms of the Fresnel images of the periodically phase-modulated input light. In particular, we demonstrate flat-top-pulse generation with a duty ratio of 50% at a quarter of the Talbot condition for the sinusoidal phase modulation. Finally, we propose a binary modulation format to generate a well-defined square-wave-type optical bit pattern.

## 1. Introduction

Generation of ultrashort optical pulses at high repetition rates is a subject of increasing interest, which finds substantial application in ultrahigh-speed optical communications [1]. Ultrashort pulses obtained directly from passively mode-locked lasers suffer from the lack of electrical control of the pulse parameters, such as pulse width, pulse shape, and pulse position in a time slot. Moreover, it is not possible to tune the repetition rate for synchronization with other electrical signals. The above limitations can be overcome by the use of external modulators that permit ultrashort pulse generation from the continuous wave (CW) light emerging from a narrowband laser. Amplitude modulators suffer from large insertion losses and a low signal-to-noise ratio [2-4]. Alternatively, phase modulators have been widely employed for pulse pattern generation [5-8]. The quasi-velocity-matched guided-wave electrooptic modulator has allowed the design of compact, stable, and low-power ultrashort optical pulse generators [9]. In a different context, Sato has demonstrated optical pulse generation from a Fabry-Perot (FP) laser [10]. Here, no external modulation is employed. The physical mechanisms involved are the gain nonlinearities and the four-wave mixing process that originate the competition among the longitudinal modes supported by the laser cavity [11-14]. The CW light emerging from the FP laser is periodically phase-modulated with a frequency that is exactly the free spectral range (FSR) of the cavity.

The electrooptic method for optical pulse generation is based on the phase modulation with a sinusoidal signal of a CW beam from a narrowband laser diode. This produces harmonic sidebands (THz) around the optical carrier frequency so that the emerging waveform is strongly chirped. The optical field is launched through a group-delay-dispersion (GDD) circuit and compressed because the sweep rate acquired upon propagation partially compensates for the chirp. Among others, a single mode optical fiber (SMF) of adjusted length, a pair of diffraction gratings, an optical synthesizer, or a linearly chirped fiber Bragg grating (LCFG) have been employed as dispersive delay lines. Specifically we mention generation of optical pulses with a temporal duration of 4.4 *ps* and with a duty ratio of 11% by means of a LCFG and an electrooptic modulator (EOM) [15].

Up to present, only an heuristic explanation for the frequency modulation (FM) to amplitude-modulation (AM) conversion process is available. The bunching parameter *B*, defined essentially as the product between the frequency chirping rate and the GDD coefficient, provides a rough estimation for the optimum bunching of the frequency components. The case of $B=1$ gives the condition under which the CW light is optimally compressed. This method shows a low pulse extinction ratio. In fact, the optical frequency of the sinusoidally phase-modulated light is assumed to be linearly chirped within half a period. Nonlinear chirped frequency components yield other substructures or broad wings, so that a considerable part of the energy lies outside the main pulse. On the other hand, note that blue-chirping and red-chirping regions are repeated in every modulation period. As a result, both the normal GDD and the anomalous one are effective for this method. The normal dispersion corresponds to compression of red-chirped portions of the input field, whereas blue-chirped portions are compressed by an anomalous dispersion circuit. Therefore, approximately half of the energy in the input field does not contribute to the bunching and generates an undesirable dc floor level. Some attempts have been done in the past few years for highly extinctive electrooptic pulse pattern generation [16]. Apart from short pulse generation with a low duty ratio, flat-top-pulse generation with a duty ratio of nearly 50% has been very recently reported [15,17]. The condition needed to generate this waveform remains unknown. These pulses can be used for instance for return-to-zero (RZ) modulation formats in optical fiber communication [18].

Here, we face electrooptic pulse pattern generation from a radically different point of view, which allows nearly background-free picosecond pulsation. We recognize that, as a result of the periodic nature of the phase modulation, when the chirped light has evolved through the GDD circuit, the output intensity is also periodic in the time coordinate, with a fundamental frequency that is, in general, the same as the one for the phase modulation *f*. But the output intensity is also periodic with the GDD coefficient $\Phi_2$. The period is given by the so-called temporal Talbot dispersion relationship, $\Phi_{2T} = 1/\pi f^2$ [19-21]. Furthermore, we show that a remarkably simple formula describes the optical intensity at a quarter of the Talbot dispersion. On the framework of the space-time analogy [22], the above results constitute the temporal analogue of the field diffracted by a pure phase grating [23-25]. The parameters of the electrooptical modulator, the frequency of the driving signal and the modulation index, or alternatively the FSR in a FP laser, together with the GDD coefficient determine unambiguously the waveform achieved at the output. Specifically, we show flat-top-pulse generation with a duty ratio of 50% for a modulation index of $\pi/4$ providing the sought theoretical support of the experimental results reported in [15,17]. The present description permits to identify a great variety of other pulse profiles. If we change continuously the dispersion amount in the GDD circuit, Fresnel patterns in intensity corresponding to a 1D sinusoidal phase-only grating appear, but now in the time domain, subsequently at the output of the arrangement. Of course, the same conclusion applies for other nonsinusoidal phase-only modulations.

This paper is structured as follows. In Section 2, the evolution of the optical field associated with a periodically phase-modulated input light through an arbitrary GDD circuit is provided in terms of the Talbot dispersion amount. We illustrate several examples concerning synthesis of different pulse waveforms at different dispersion amounts. In Section 3, the output pulse intensity is expressed in terms of a simple trigonometric formula when the output dispersion corresponds to a quarter of the Talbot dispersion. We identify an ultra-flat-top-pulse pattern by binary phase modulation of CW light. Finally, in Section 4, the effect of the third order dispersion (TOD) of the SMF, or alternatively the spectral window of a LCFG, when used as a GDD circuit is discussed.

## 2. Theoretical analysis

After phase modulation, the optical field of the narrowband CW light is expressed as

$$E_{in}(t) = E_o \exp(-j\omega_o t)\exp[jV(t)] \quad . \tag{1}$$

Here $E_o$ is the constant amplitude, $\omega_o$ denotes the carrier optical frequency, and $V(t)$ is the phase modulation function. For our purposes, we assume that $V(t)$ is a periodic function with period $T$. Note that the perfect sinusoid is enclosed as a particular case. As a result of the periodicity of the phase $V(t)$, we can rewrite Eq. 1 in terms of a Fourier series expansion, namely,

$$E_{in}(t) = E_o \exp(-j\omega_o t)\sum_{n=-\infty}^{\infty} c_n \exp\left(-j2\pi n\frac{t}{T}\right) \quad . \tag{2}$$

The periodic optical input intensity is

$$I_{in}(t) = |E_{in}(t)|^2 = |E_o|^2 \sum_{N=-\infty}^{\infty} C_N \exp\left(-j2\pi N\frac{t}{T}\right) \quad , \tag{3}$$

where

$$C_N = \sum_{n=-\infty}^{\infty} c_n^* c_{n+N} \quad . \tag{4}$$

Of course, from Eq. 1, $I_{in}(t) = |E_o|^2$. This implies that $C_N = \delta_{N,0}$, where $\delta_{N,0}$ denotes the Kronecker delta function.

Aside from an irrelevant constant factor, the phase delay of an ideal GDD circuit is

$$H(\omega) = \exp[j\Phi_1(\omega - \omega_o)]\exp[j\Phi_2/2(\omega - \omega_o)^2] \quad , \tag{5}$$

with $\Phi_1$ and $\Phi_2$ denoting the group delay and the GDD coefficient, respectively. Note that we assume no losses neither in the coupling of the input into the dispersive circuit or in the propagation. If we consider that the GDD circuit is implemented using a SMF, $\Phi_1 = \beta_1 z$ and $\Phi_2 = \beta_2 z$, with $z$ the propagation distance. The parameters $\beta_1$ and $\beta_2$ are the inverse of the group velocity and the group velocity dispersion (GVD) parameter of the fiber, respectively.

Note that we neglect higher-order dispersion terms and nonlinear interactions. Roughly speaking, both assumptions are satisfied when the bandwidth of the input light is less than $3|\beta_2/\beta_3|$, with $\beta_3$ the TOD parameter of the fiber, and the power carried by individual pulses is not enough to excite nonlinear mechanisms in the fiber [26]. In section 4 we will further consider the narrowband assumption.

After propagation inside the GDD circuit (see Fig.1) the output field becomes

$$E_{out}(t,\Phi_2) = E_o \exp(-j\omega_o t)\sum_{n=-\infty}^{\infty} c_n \exp\left(j\frac{\Phi_2 4\pi^2 n^2}{2T^2}\right)\exp\left(-j2\pi n\frac{(t-\Phi_1)}{T}\right) \quad . \tag{6}$$

From now on, the description of the signal is given in a reference framework moving at the group velocity of the wave packet, i.e., $\tau = t - \Phi_1$. From Eq. 6, the output intensity can be written as

$$I_{out}(\tau,\Phi_2) = |E_{out}(\tau,\Phi_2)|^2 = |E_o|^2 \sum_{N=-\infty}^{\infty} C'_N(\Phi_2)\exp\left(-j2\pi N\frac{\tau}{T}\right) , \qquad (7)$$

with

$$C'_N(\Phi_2) = \exp\left(j\frac{\Phi_2}{\Phi_{2T}}2\pi N^2\right)\sum_{n=-\infty}^{\infty} c_n^* c_{n+N} \exp\left(j\frac{2\Phi_2}{\Phi_{2T}}2\pi nN\right) . \qquad (8)$$

Two findings are clear from the above equations. First, Eq. 7 indicates that $I_{out}(\tau,\Phi_2)$ is a periodic function of $\tau$. Its period is, in principle, equal to the modulation period $T$. Second, from Eqs. 7 and 8 it is clear that the output optical intensity changes periodically with the dispersion coefficient $\Phi_2$. The period is just the Talbot dispersion, $\Phi_{2T} = 1/\pi f^2$. From Eq. 8 we note that $C'_N(\Phi_2 + \Phi_{2T}/2) = \exp(j\pi N^2)C'_N(\Phi_2)$. In this way, we obtain $I_{out}(\tau,\Phi_2) = I_{out}(\tau + T/2, \Phi_2 + \Phi_{2T}/2)$. This means that a change in the dispersion by $\Phi_{2T}/2$ is equivalent to a temporal shift of half a period at the output intensity. We explore further implications of the above facts.

Next, we consider, as an example, the case of perfect sinusoidal modulation, $V(t) = \Delta\theta \sin(2\pi f t)$. Here, $\Delta\theta$ is the modulation index in radians. Of course $f = 1/T$. For this case, the Fourier coefficients are expressed by the Bessel functions of the first kind, $c_n = J_n(\Delta\theta)$. Therefore,

$$C'_N(\Phi_2) = \exp\left(j\frac{\Phi_2}{\Phi_{2T}}2\pi N^2\right)\sum_{n=-\infty}^{\infty} J_n(\Delta\theta) J_{n+N}(\Delta\theta) \exp\left(j\frac{2\Phi_2}{\Phi_{2T}}2\pi nN\right) . \qquad (9)$$

To illustrate waveform formation, we consider a realistic example concerning sinusoidal modulation where $\Phi_2$ and $f$ are set to $\Phi_2 = \Phi_{2T}/16$ and $40$ GHz, respectively. Different new pulse waveforms not yet reported are obtained by changing $\Delta\theta$. In particular, we mention short pulse generation for $\Delta\theta = \pi/4$. Here, a duty cycle (DC) of 33% is achieved. Note that in this case the signal is free of annoying wings and tails but a high dc-floor level is present, as shown in Fig. 2(a). In Fig. 2(b) the numerical simulation shows a short pulse with a DC of approximately 18%. Although part of the energy lies outside the main pulse, the remaining dc-floor level is low. Note that, aside for a temporal shift of half a period, the same profiles are achieved when the dispersion is set to $\Phi_2 = \Phi_{2T}(8q+1)/16$ where $q$ is an arbitrary integer. We also claim that the above shapes can be achieved with normal GDD as well as with anomalous one.

## 3. Flat-top-pulse generation

In this section we particularize the above key equations when dispersion is set to a quarter of the Talbot dispersion. From Eqs. 7 and 8, for $\Phi_2 = \Phi_{2T}/4$ we obtain (see Appendix)

$$I_{out}(\tau, \Phi_2 = \Phi_{2T}/4) = |E_o|^2\{1 - \sin[V(\tau - T/2) - V(\tau)]\} . \qquad (10)$$

Equation 10, which is one of the main results of this paper, provides theoretical support for electrooptic flat-top-pulse generation, as will be shown next. At this point it is worth mentioning that the spatial analogue of the above formula was derived in [24,25], in the context of Fourier optics, to describe the properties of the irradiance distribution corresponding to the Fresnel diffraction patterns of a one-dimensional phase grating. Therefore, one should anticipate the above result within the framework of the celebrated space-time analogy.

We note that Eq. 10 is valid for a general periodic phase function $V(\tau)$. If we consider the sinusoidal modulation $V(\tau) = \Delta\theta \sin(2\pi f \tau)$, then

$$I_{out}(\tau, \Phi_2 = \Phi_{2T}/4) = |E_o|^2 \{1 + \sin[2\Delta\theta \sin(2\pi f \tau)]\} \quad . \tag{11}$$

From Eq. 11 we observe that when $\Delta\theta = \pi/4$, the argument within the exterior trigonometric function ranges from $-\pi/2$ to $\pi/2$ for $\tau \in [-T/2, T/2]$. The analytical curve shown in Eq. 11 is plotted in Fig. 3 for $\Delta\theta = \pi/4$ and an input frequency of *40* GHz. The temporal width of the individual pulses is *12.5* ps. In this way, a nearly flat-top-pulse with a DC of 50% is achieved. Equation 11 provides an analytical formula for the waveform that was experimentally obtained in references [15] and [17]. Furthermore, due to the periodic nature of the optical field, the same result is achieved for a GDD dispersion $\Phi_2 = \Phi_{2T}(2q+1)/4$, with $q$ an arbitrary integer. The existence of multiple GDD amounts was pointed out in [15].

Next, we seek a different phase modulation format that allows ultra-flat-top-pulse generation. With this aim, we consider the periodic binary phase-only modulation of the carrier frequency given by

$$V(\tau) = \begin{cases} 0 & \text{if } \tau \in [0, T/2) \\ \pi/2 & \text{if } \tau \in [T/2, T) \end{cases}, \tag{12}$$

with *T* being the period. The modulation $V(\tau)$ is plotted in Fig. 4(a). For this case, the argument inside the trigonometric function in Eq. 10 has two values, $-\pi/2$ and $\pi/2$, respectively. Consequently, the output intensity shows a binary shape at $\Phi_2 = \Phi_{2T}/4$, namely

$$I_{out}(\tau, \Phi_2 = \Phi_{2T}/4) = \begin{cases} 0 & \text{if } \tau \in [0, T/2) \\ 2|E_o|^2 & \text{if } \tau \in [T/2, T) \end{cases}. \tag{13}$$

In order to clarify our description, we have calculated numerically and plotted in Fig. 4(b), $I_{out}(\tau, \Phi_2)$ for the phase modulation in Eq. 12 and $\Phi_2$ ranging the whole first Talbot period. As expected, for dispersions $\Phi_2 = 0$, $\Phi_2 = \Phi_{2T}/2$, and $\Phi_2 = \Phi_{2T}$, the irradiance presents a constant value. Whereas for $\Phi_2 = \Phi_{2T}/4$, according to Eq. 13 an ultra-flat-top optical pulse train is obtained, see Fig. 4(c). This kind of pulse could be employed for RZ modulation formats in optical signal transmission and, in particular, for differential phase-shift-keyed transmission.

### 4. GDD circuit analysis

A. Standard SMF

It is usual to perform the FM to AM conversion process by means of a SMF. The strongly chirped light emerging from the electrooptical modulator is temporally distorted and compressed due to the propagation inside the fiber. A rigorous analysis of the quadratic approximation in Eq. 5 must be carried out to test the performance of the setup. Generally speaking, the spectral bandwidth of the incoming signal, $\Delta\omega$, should be limited to $\Delta\omega < 3|\beta_2/\beta_3|$. For the case of perfect sinusoidal modulation, we have $c_n = J_n(\Delta\theta)$. To obtain a rough estimation for the optical bandwidth of the phase-modulated signal, we plot in Fig. 5 $J_n(\Delta\theta)$ versus the modulation index $\Delta\theta$. Four different values of the order *n* have been considered. The modulation index ranges within the interval $0 < \Delta\theta < 10$. From this plot we can assume that the main contribution to the output intensity comes from the Bessel functions with an order lower than 10. Thus, the condition for the validity of the parabolic approximation reads $20 f < 3|\beta_2/\beta_3|$. If we have an optical source peaked at the $1.55\,\mu m$ window, for a standard SMF we obtain $\beta_2 = -2.168 \times 10^{-2} \, ps^2/m$ and $\beta_3 = 1.2661 \times 10^{-4} \, ps^3/m$. In this way, the value of the term $(3/20)|\beta_2/\beta_3|$ is approximately *25THz*. So, the above inequality is widely satisfied even for the fastest commercially available electrooptical modulator, which works in the *GHz* range.

### B. LCFG

The response of a LCFG operating in reflection is assumed to be a phase quadratic function only over a limited bandwidth $\Delta\Omega$. As a result two conditions must be fulfilled for the use of a LCFG as dispersive element. First, the carrier frequency should match the central frequency of the reflected spectral band of the grating. Second, the full spectral bandwidth of the phase-modulated light, $\Delta\omega$, must be lower than $\Delta\Omega$. The spectral bandwidth of the element is related to the length through the expression $L = c|\Phi_2|\Delta\Omega/2n_{eff}$ [27], with $n_{eff}$ the effective refractive index. We assume that the LCFG is designed to match the condition $\Phi_2 = \Phi_{2T}/4$ and, analogous to the previous case, $\Delta\omega$ can be roughly estimated to be *20f*. Thus, the device will provide a linear time delay whenever $Lf > 20 \, c/8 n_{eff}$. A LCFG *4 cm* long is good enough for $f = 25 \, GHz$.

### 5. Conclusions

The evolution of an input field, consisted in a periodically phase-modulated CW light, into an arbitrary GDD circuit has been carried out in terms of the Talbot dispersion. The periodicity of the phase function has allowed us to derive an analytical formula for the FM to AM conversion process at one quarter of the Talbot dispersion. Furthermore, we have provided theoretical support for the recently experimentally demonstrated generation of a flat-top-pulse train using a phase modulator and a LCFG. We have also considered the generation of ultra-flat-top light pulses by phase modulation with a square-wave-type signal, which could be employed for RZ modulation format. We numerically show that a SMF or LCFG can be used as an efficient GDD device in terms of the spectral bandwidth of the modulated signal. We would like to mention that the mathematical framework developed in this work is also applicable to the case of a multimode laser source, such as a FP laser, working in the frequency-modulated supermode regime. In this case, the frequency of the equivalent modulator will be given by the free spectral range of the longitudinal modes supported by the laser cavity. In the framework of the space-time analogy, the above results constitute the temporal analogue of the Fresnel diffraction field diffracted by a pure phase grating.

### Appendix

We note that the Fourier series expansion of $\exp(jV(\tau))$ in Eq. 2 can be rewritten as

$$M(\tau) = \exp(jV(\tau)) = \sum_{n=-\infty}^{\infty} c_{2n} \exp\left(-j2\pi(2n)\frac{\tau}{T}\right) + \\ + \sum_{n=-\infty}^{\infty} c_{2n+1} \exp\left(-j2\pi(2n+1)\frac{\tau}{T}\right) \quad . \tag{A1}$$

From the above equation, we have

$$M(\tau) - M(\tau - T/2) = 2\sum_{n=-\infty}^{\infty} c_{2n+1} \exp\left(-j2\pi(2n+1)\frac{\tau}{T}\right) \quad , \tag{A2}$$

$$M(\tau) + M(\tau - T/2) = 2\sum_{n=-\infty}^{\infty} c_{2n} \exp\left(-j2\pi(2n)\frac{\tau}{T}\right) \quad , \tag{A3}$$

and, consequently,

$$[M(\tau)-M(\tau-T/2)][M^*(\tau)+M^*(\tau-T/2)]=2j\sin[V(\tau-T/2)-V(\tau)] \quad . \tag{A4}$$

Let us now particularize Eq. 8 for $\Phi_2 = \Phi_{2T}/4$. We obtain

$$I_{out}(\tau,\Phi_2=\Phi_{2T}/4)=|E_o|^2\left\{\sum_{N=-\infty}^{\infty}C'_{2N+1}(\Phi_{2T}/4)\exp\left(-j2\pi(2N+1)\frac{\tau}{T}\right)+\right.$$
$$\left.+\sum_{N=-\infty}^{\infty}C'_{2N}(\Phi_{2T}/4)\exp\left(-j2\pi(2N)\frac{\tau}{T}\right)\right\} \quad , \tag{A5}$$

where, taking into account Eq. 4,

$$C'_{2N}(\Phi_{2T}/4)=\sum_{n=-\infty}^{\infty}c_n^*c_{n+2N}=\delta_{2N,0} \quad , \tag{A6}$$

$$C'_{2N+1}(\Phi_{2T}/4)=2j\sum_{n=-\infty}^{\infty}c_{2n}^*c_{2n+2N+1} \quad . \tag{A7}$$

Substitution of Eqs. A6 and A7 into Eq. A5 after some simple algebra leads to

$$I_{out}(\tau,\Phi_2=\Phi_{2T}/4)=|E_o|^2\{1-\sin[V(\tau-T/2)-V(\tau)]\} \quad , \tag{A8}$$

which is Eq. 10 in the text.

### Acknowledgments

The authors gratefully acknowledge fruitful discussions with Professor Jorge Ojeda-Castañeda and Dr. Juan Carlos Barreiro about array illuminators and Talbot effect, which inspired this work. This research was funded by the *Dirección General de Investigación Científica y Técnica*, Spain, under the project FIS2004-02404. Partial financial support from projects TEC2004-04754-C03-02 and UNJM-E025, *fondos* FEDER-MCT, is also acknowledged. Víctor Torres gratefully acknowledges financial assistance from a FPU grant of the *Ministerio de Educación y Ciencia*, Spain.

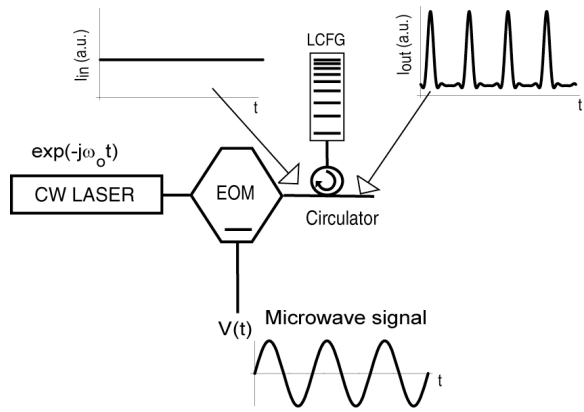

Figure 1

Schematic diagram of the electrooptic pulse generator. The GDD circuit is implemented by means of a LCFG working in reflection.

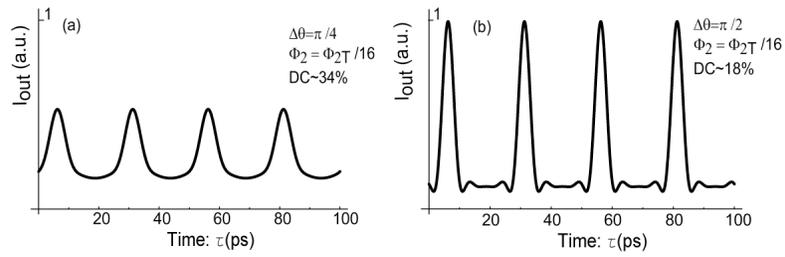

Figure 2

Simulated pulse waveforms obtained by dispersion of sinusoidally phase-modulated light through a GDD circuit.

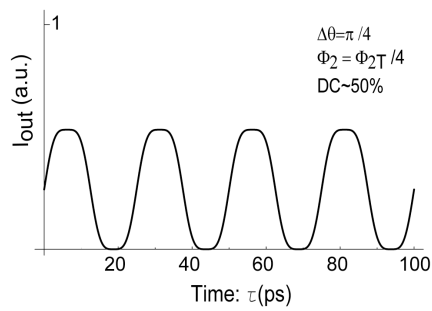

Figure 3

Flat-top-pulse generation under sinusoidal phase modulation.

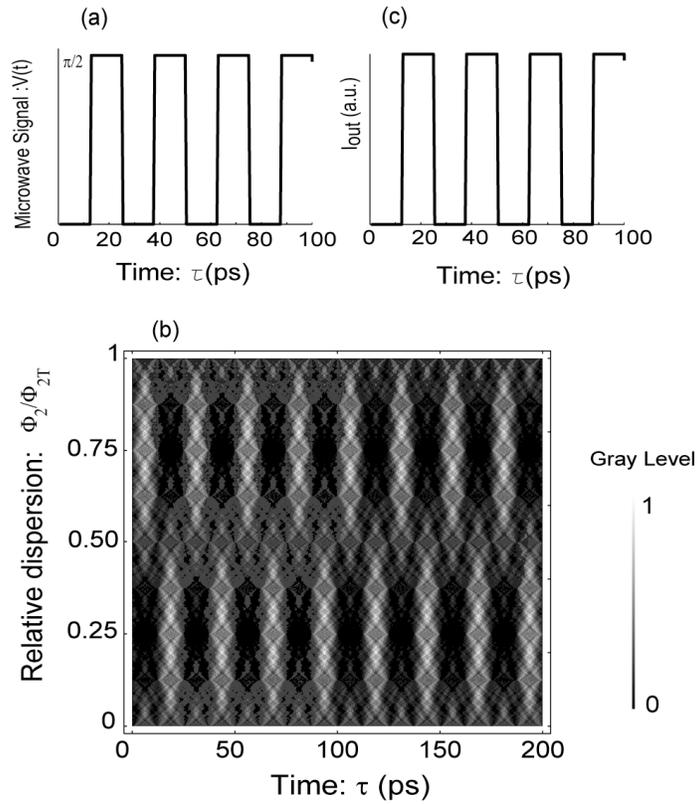

Figure 4

Ultra-flat top, RZ-format-pulse generation: **a)** phase modulation function $V(t)$; **b)** Numerically evaluated output intensity at dispersions values ranging the interval $0 \leq \Phi_2/\Phi_{2T} \leq 1$; and **c)** Plot of $I_{out}(\tau, \Phi_2/\Phi_{2T} = 0.25)$ given by Eq. 13 for $f = 40 GHz$.

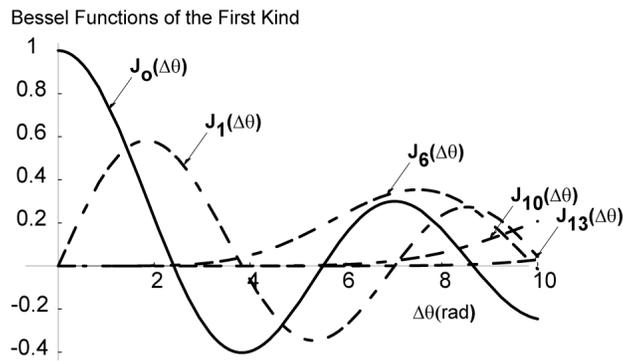

Figure 5

$J_n$ versus the modulation index $\Delta\theta$ for four different values of *n*.